\documentclass[times,twocolumn,final,longtitle]{elsarticle}

\usepackage{medima_arxiv}
\usepackage{framed,multirow}
\usepackage{booktabs} 
\usepackage{amssymb}
\usepackage{latexsym}
\usepackage{hyperref}
\usepackage{amsmath}
\usepackage{enumitem}
\usepackage{multirow}
\usepackage{makecell}
\usepackage[german, english]{babel}
\usepackage{url}
\usepackage{xcolor}

\usepackage{hyperref}

\usepackage{algorithm}
\usepackage{algpseudocode}
\usepackage{float}

\definecolor{newcolor}{rgb}{.8,.349,.1}

\journal{Medical Image Analysis}

\begin{document}

\verso{Ruijie Yang \textit{et~al.}}

\begin{frontmatter}

\title{EndoFinder: Online Lesion Retrieval for Explainable Colorectal Polyp Diagnosis Leveraging Latent Scene Representations}
\communicated{}
\author[1,2,3,4]{Ruijie \snm{Yang}\fnref{fn1}}
\fntext[fn1]{These authors contribute equally.}
\author[5,6]{Yan \snm{Zhu}\fnref{fn1}}
\author[5,6]{Peiyao \snm{Fu}}
\author[7]{Yizhe \snm{Zhang}}
\author[4]{Zhihua \snm{Wang}\corref{cor1}}
\cortext[cor1]{Corresponding author.}
\ead{zhihua.wang@zju.edu.cn }
\author[5,6]{Quanlin \snm{Li}}
\author[5,6]{Pinghong \snm{Zhou}}
\author[8]{Xian \snm{Yang}}
\author[1,3]{Shuo \snm{Wang}\corref{cor1}}
\ead{shuowang@fudan.edu.cn }

\address[1]{Digital Medical Research Center, School of Basic Medical Sciences, Fudan University, Shanghai, China}
\address[2]{School of Software Technology, Zhejiang University, Ningbo, Zhejiang, China}
\address[3]{Shanghai Key Laboratory of MICCAI, Shanghai, China}
\address[4]{Shanghai Institute for Advanced Study of Zhejiang University, Shanghai, China}
\address[5]{Endoscopy Center and Endoscopy Research Institute, Zhongshan Hospital, Fudan University, Shanghai, China}
\address[6]{Shanghai Collaborative Innovation Center of Endoscopy, Shanghai, China}
\address[7]{School of Computer Science and Engineering, Nanjing University of Science and Technology, Nanjing, Jiangsu, China}
\address[8]{Alliance Manchester Business School, The University of Manchester, Manchester, UK}

\begin{abstract}
Colorectal cancer (CRC) remains a leading cause of cancer-related mortality, underscoring the importance of timely polyp detection and diagnosis. While deep learning models have improved optical-assisted diagnostics, they often demand extensive labeled datasets and yield “black-box” outputs with limited interpretability. In this paper, we propose EndoFinder, an online polyp retrieval framework that leverages multi-view scene representations for explainable and scalable CRC diagnosis.
First, we develop a Polyp-aware Image Encoder by combining contrastive learning and a reconstruction task, guided by polyp segmentation masks. This self-supervised approach captures robust features without relying on large-scale annotated data. Next, we treat each polyp as a three-dimensional “scene” and introduce a Scene Representation Transformer, which fuses multiple views of the polyp into a single latent representation. By discretizing this representation through a hashing layer, EndoFinder enables real-time retrieval from a compiled database of historical polyp cases, where diagnostic information serves as interpretable references for new queries.
We evaluate EndoFinder on both public and newly collected polyp datasets for re-identification and pathology classification. Results show that EndoFinder outperforms existing methods in accuracy while providing transparent, retrieval-based insights for clinical decision-making. By contributing a novel dataset and a scalable, explainable framework, our work addresses key challenges in polyp diagnosis and offers a promising direction for more efficient AI-driven colonoscopy workflows. The source code is available at https://github.com/ku262/EndoFinder-Scene.
\end{abstract}


\end{frontmatter}


\section{Introduction}
Colorectal cancer (CRC) represents a significant public health challenge, ranking as the second leading cause of cancer-related mortality overall and the third most commonly diagnosed malignancy~\citep{Siegel2024cancer}. Early detection, diagnosis, and treatment are crucial for reducing CRC incidence and mortality rates~\citep{van2006polyp}. Colonoscopy is recognized as the gold standard screening procedure for both preventing and detecting CRC at an early stage, primarily through the identification and management of colorectal polyps~\citep{pamudurthy2020advances}. During these examinations, endoscopists are faced with critical intraprocedural decisions regarding polyp management, specifically whether to resect potentially malignant lesions or to surveil those presumed to be benign~\citep{chandran2015can}. While histopathological analysis of biopsied or resected tissue remains the definitive diagnostic standard, these results are not available in real-time during the procedure. Consequently, clinicians must often rely on their experience and subjective visual assessment to make immediate decisions, particularly for small colorectal polyps. This reliance introduces inherent variability and potential limitations into clinical practice.

To address this challenge and enhance clinical decision-making during colonoscopy, artificial intelligence (AI)-driven optical biopsy has emerged as a promising approach~\citep{ribeiro2016colonic}. Typically, these AI systems are developed by training deep learning algorithms on datasets comprising endoscopic images paired with their corresponding histopathological outcomes~\citep{ribeiro2016exploring, korbar2017deep}. Such systems have demonstrated capabilities comparable to experienced physicians, particularly in the classification of small polyps~\citep{younas2023deep, komeda2017computer}. By facilitating reliable, real-time classification of these diminutive lesions, AI offers the potential to significantly reduce unnecessary resections of non-neoplastic polyps, thereby mitigating overtreatment risks and alleviating the burden on patients and healthcare resources~\citep{barua2022real}. However, the predominant supervised learning approaches used to train these models face significant hurdles. They demand extensive annotated datasets which are costly and labor-intensive to produce~\citep{kundu2020exploiting}. Furthermore, as datasets expand, these inductive models often require complete retraining to assimilate new information, presenting considerable scalability challenges. Moreover, these models often exhibit a 'black box' nature, generating diagnostic results without clear explanatory pathways~\citep{apley2020visualizing}, hindering clinical acceptance and trust among physicians.

Motivated by these limitations, we propose a novel transductive framework centered on self-supervised learning and retrieval-based diagnostics. When a new polyp is encountered, we extract its visual features and retrieve the most similar polyps (i.e., digital twins) from a pre-compiled database of historical cases. The diagnosis for the new polyp is then inferred from the known pathological reports of these retrieved similar cases, providing inherent interpretability by example. To enable efficient real-time application during colonoscopy, we discretize the polyp representations into binary features and construct a ball tree~\citep{brearley2022knn} based on Hamming distance~\citep{cao2018deep}.  This optimization significantly accelerates the search for similar polyps, making near-instantaneous retrieval feasible in a clinical setting.


Despite supervised classifiers or transductive approaches, effective representation learning is crucial in medical image analysis~\citep{liao2013representation}. For colorectal polyps, representation learning is particularly challenging because their color and texture often closely resemble the surrounding colonic tissue~\citep{hsu2021colorectal}, leading to subtler visual distinctions compared to those found in natural images~\citep{li2021qhash}. Furthermore, polyps are inherently three-dimensional structures, making it impossible to capture a fully comprehensive representation from a single two-dimensional endoscopic image. To overcome the challenges of learning robust polyp representations, we initially developed the polyp-aware image encoder, a specialized feature extractor that focuses on the relevant regions of polyp~\citep{yang2024endofinder}. However, recognizing that a single 2D view is inherently limited due to the three-dimensional nature of polyps and the multi-angle perspectives typically captured during colonoscopy~\citep{shi2006ct}, we sought to further enhance the representation. Inspired by recent progress in computer vision for modeling complex 3D structures, we propose conceptualizing the polyp as a 'scene'. This paradigm shift draws from advancements like Neural Radiance Fields (NeRF)~\citep{mildenhall2021nerf}, which excel at synthesizing novel views but demand precise camera poses and extensive computation. Subsequent methods like the Scene Representation Transformer (SRT)~\citep{sajjadi2022scene} offered improvements by encoding multiple views into a generalizable latent representation, although still requiring posed inputs. More aligned with the unconstrained nature of colonoscopic acquisition, the Really Unposed Scene Representation Transformer (RUST)~\citep{sajjadi2023rust} demonstrated the potential for learning powerful scene representations directly from unposed image collections, enabling tasks like novel view synthesis without explicit pose data. 

Building upon these concepts, we propose a scene representation transformer to process 2D polyp features extracted from various viewing angles, integrating this multi-view information to create a richer, more holistic representation that implicitly captures 3D characteristics. This multi-view approach aims to provide a more comprehensive and clinically relevant analysis compared to single-image methods. In summary, our key contributions are as follows: 
\begin{itemize}
    \item We introduce and release PolypScene-250, a novel multi-view polyp dataset, including the PolypScene-80 subset with histopathology annotation, to advance polyp characterization research.
    \item We propose a holistic self-supervised strategy using contrastive learning and reconstruction for both a polyp-aware image encoder and a multi-view scene encoder.
    \item We develop EndoFinder, a retrieval-based diagnostic framework that uses scene representations, hashing, and database search to provide interpretable diagnoses based on similar historical cases.
    \item We demonstrate through re-identification and classification experiments that our method learns robust representations of polyps and provides valuable, interpretable diagnostic references to support clinical decision-making.
\end{itemize}

A preliminary version of this work was published at MICCAI 2024~\citep{yang2024endofinder}. This journal submission significantly extends that work by moving from 2D image retrieval to a 2.5D/3D scene retrieval paradigm. We achieve this by explicitly modeling polyps as scenes and leveraging multi-view perspectives to capture implicit 3D information within a latent scene representation.  We highlight some key improvements: (1) Explicitly addressing the limitation of single-image representations by incorporating multi-view inputs. (2) Designing an encoder-decoder architecture for scene representation learning, utilizing global features from multiple views to generate a robust latent representation and improving scene understanding via view reconstruction tasks.

\section{Related Work}
This section reviews prior work relevant to our approach, focusing on supervised methods for optical polyp diagnosis, the use of content-based image retrieval in medical image analysis, and recent developments in 3D scene representation learning.

\begin{figure*}[!t]
\centering
\includegraphics[scale=.25]{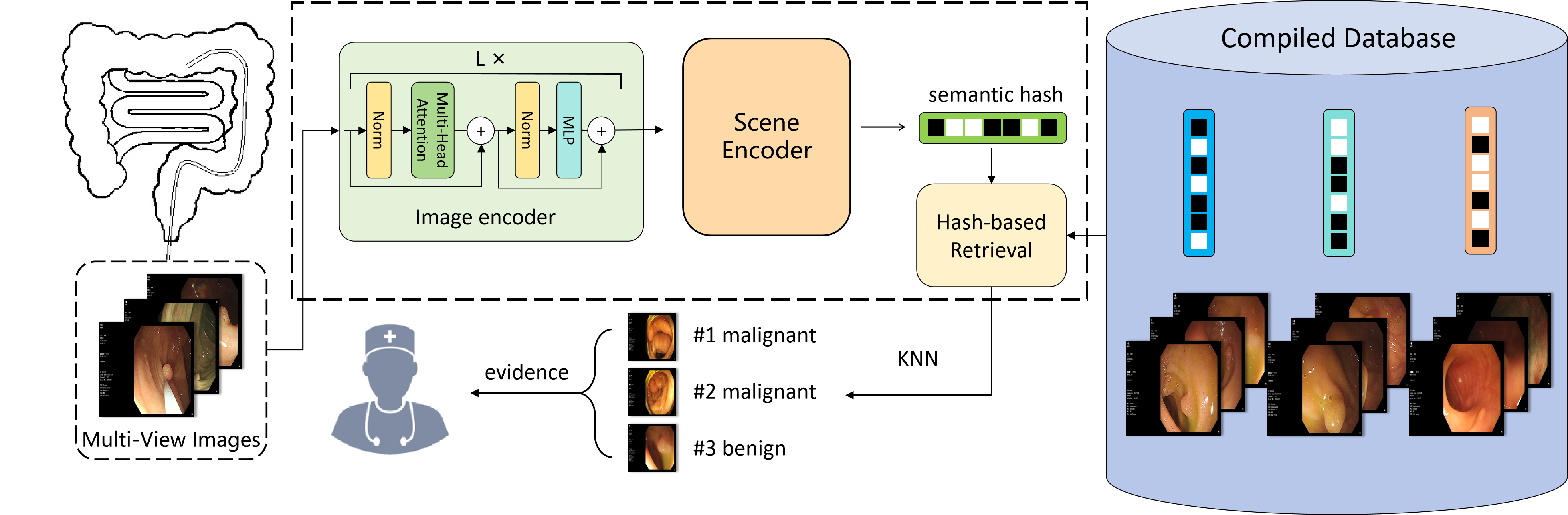}
\caption{Overview of the proposed EndoFinder framework. Endoscopic multi-view images are encoded into latent scene representations and discretized into hash codes for fast lesion retrieval. The decision-making is augmented by referring to the historical information of similar polyps in the database.}
\label{fig:workflow}
\end{figure*}

\subsection{Computer-Assisted Polyp Diagnosis}
Various methods and models have been proposed for polyp classification in colonoscopy, which can be broadly categorized into two types: traditional computer vision methods and deep learning-based methods. In traditional computer vision, the LFD (Local Fractal Dimension) method was applied to colorectal polyp classification~\citep{hafner2015local}. This approach effectively enhanced discriminative power by extracting shape or gradient information from images. To achieve rotational invariance, the max-min filter bank strategy was employed to distinguish between different polyps~\citep{wimmer2016novel}. Additionally, a combination of wavelet transformations and Weibull features achieved improved results~\citep{wimmer2016directional}. With the remarkable performance of convolutional neural networks (CNNs) in image classification, CNN-based systems have been utilized to improve the accuracy of small colorectal polyp assessments~\citep{jin2020improved}, thereby reducing the diagnostic time for physicians. Various convolutional models, such as VGG, ResNet, DenseNet, SENet, and MnasNet, have been used for polyp classification, and their performance has been compared~\citep{patel2020comparative}. Results indicated that modern CNN models could successfully classify polyps with accuracy comparable to, or even surpassing, that reported by gastroenterologists. To assist non-expert endoscopists in diagnosing colorectal tumors as accurately as expert endoscopists during colonoscopy, an AI system based on ResNet152 was developed~\citep{yamada2022robust}. This system automatically and robustly predicted pathological diagnoses according to the Revised Vienna Classification, consistently outperforming expert endoscopists in both internal and external validations. Additionally, to automate the classification of polyps according to the Paris Classification, a state-of-the-art accuracy was achieved by cropping and segmenting the polyps before classification using a Vision Transformer (ViT)~\citep{krenzer2023automated}. Despite the impressive progress and high classification accuracy achieved by deep learning-based methods, these approaches still face critical limitations, particularly in terms of interpretability and clinical transparency. Most state-of-the-art models operate as "black boxes," providing predictions without clear explanations, which hinders trust and widespread adoption in real-world clinical settings. As a result, enhancing model explainability and aligning AI decisions with medical reasoning remain essential research directions for the safe and effective deployment of AI-assisted diagnostic systems in colonoscopy. 

\subsection{Content-based Image Retrieval for Medical Analysis}
To enhance transparency and assist decision-making, content-based image retrieval (CBIR) has gained attention as a complementary strategy. By retrieving visually or semantically similar cases from large-scale archives, CBIR offers clinicians intuitive reference points for diagnosis, training, and comparative study. In medical imaging, CBIR is particularly effective in scenarios where large quantities of unlabeled data are available. For instance, whole-slide image (WSI) retrieval has emerged as a promising tool in pathology, enabling retrieval of highly relevant diagnostic cases from historical WSI archives~\citep{wang2023retccl}. A core challenge in CBIR is the construction of a robust embedding space and an efficient search mechanism. In the field of natural images, methods like SSCD~\citep{pizzi2022self} apply SimCLR-based contrastive learning and entropy regularization to improve the separability of descriptor vectors, significantly enhancing copy detection accuracy. Transformer-based architectures such as Image Retrieval Transformer (IRT)~\citep{el2021training} further improve retrieval performance by learning more global and semantically rich representations through entropy-aware objectives. To meet the demands of real-time or large-scale retrieval, hashing techniques have been increasingly applied in medical image CBIR. CBMIR~\citep{ozturk2021class} reduces the dimensionality of CNN features and transforms them into compact hash codes for faster retrieval. Precision medical hash retrieval methods~\citep{guan2022precision} introduce a dedicated hashing layer and optimize multiple loss functions, including classification, quantization, and bit balance, to generate highly discriminative binary codes. Similarly, MTH (Multi-scale Triplet Hashing)~\citep{chen2023multi} leverages multi-scale context, convolutional self-attention, and hierarchical similarity learning to improve retrieval performance. Despite these advances, constructing generalizable representations for polyp retrieval in endoscopic imagery remains underexplored. Compared to static pathology slides, polyps exhibit higher variability in shape, texture, and illumination, posing unique challenges for retrieval-based analysis. More importantly, most existing retrieval methods rely on single-frame representations, which may fail to capture the full 3D structure of polyps that are inherently volumetric. Given the spatial complexity of endoscopic scenes, effective retrieval and understanding require models capable of reasoning over multiple views and inferring implicit 3D information.

\subsection{Scene Representation Learning}
Colorectal polyps are inherently three-dimensional structures, and capturing their 3D geometry is crucial for accurate diagnosis, classification, and treatment planning. Inferring the structure of a 3D scene from 2D images is a fundamental challenge in computer vision. NeRF (Neural Radiance Field)~\citep{mildenhall2021nerf} represents a scene using a fully connected deep network, generating novel views by projecting output colors and densities onto an image based on input spatial positions and viewing directions, using classical volume rendering techniques. However, NeRF typically requires input images with precise camera positions, and processing each new scene demands significant time. SRT (Scene Representation Transformer)~\citep{sajjadi2022scene} encodes different views of a scene into a "Set-Latent Scene Representation" and uses a given ray pose to attend to the scene representation for generating novel views. SRT can train a single model capable of providing effective latent representations that generalize beyond individual scenes but still requires accurate pose information. In an advancement of the SRT approach, RUST (Really Unposed Scene Representation Transformer)~\citep{sajjadi2023rust} enables novel view synthesis without needing precise pose information, allowing it to be applied to more diverse and larger-scale scenes. RUST achieves this by using a pose encoder to examine the target image, learning a latent pose embedding, which is then utilized by the decoder for view synthesis. In the medical domain, especially in endoscopic imaging, similar efforts have been made to enhance 3D scene understanding. Traditional monocular endoscopy inherently lacks depth perception, posing challenges for accurate diagnosis and navigation. To overcome this, researchers have developed dedicated hardware solutions such as structured-light-based endoscopic 3D scanners~\citep{schmalz2012endoscopic}, which can reconstruct dense point clouds in real time to support intraoperative visualization and measurement. On the other side, numerous algorithms have been proposed for 3D reconstruction from routine endoscopic images, including both geometry-based and learning-based methods. The extensive literature on 3D reconstruction in endoscopy, as surveyed in~\citep{yang20243d}, highlights a strong and sustained interest in integrating geometric information into endoscopic imaging workflows. Building on these trends, neural-field-based models have also been introduced to explicitly reconstruct deformable surfaces from stereo endoscopy videos. For example, EndoSurf~\citep{zha2023endosurf} models the geometry, texture, and dynamics of soft tissues via signed distance and radiance fields, achieving high-fidelity reconstructions through differentiable rendering. In parallel, representation learning from endoscopic videos has emerged as an alternative to explicit modeling. M\textsuperscript{2}CRL~\citep{hu2024multi} applies multi-view masked contrastive learning to extract spatially and temporally rich features from large-scale endoscopic video datasets, yielding strong performance across classification, segmentation, and detection tasks.  

In summary, prior work in computer-assisted polyp diagnosis has focused on classification accuracy, often neglecting interpretability. CBIR offers a path to transparency but is typically limited to single-view analysis. Meanwhile, scene representation learning shows promise for modeling 3D geometry but has not been widely integrated into medical retrieval frameworks. To bridge these gaps, we propose EndoFinder, a holistic framework that learns polyp-aware, multi-view scene representations for retrieval-based diagnosis, offering both robust performance and valuable clinical interpretability.

\section{Methods}

\begin{figure*}[h]
\centering
\includegraphics[scale=.25]{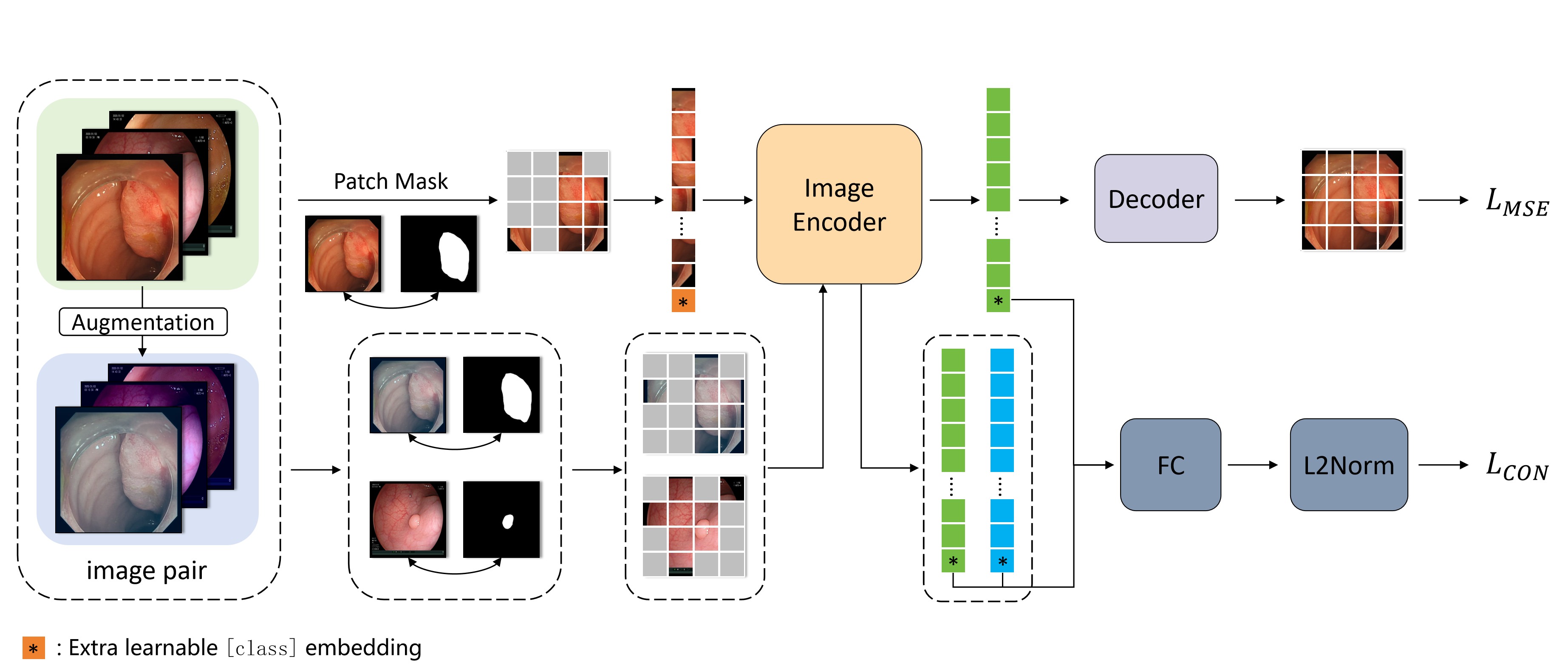}
\caption{
Overview of the polyp-aware self-supervised representation learning framework. The model jointly learns representations by contrastive learning and masked reconstruction, guided by polyp-region masks and a ViT-based image encoder. 
}
\label{fig:one-view}
\end{figure*}

\subsection{Problem Formulation}
We consider a database $D = \{(S_i, y_i)\}_{i=1}^N$, containing $N$ historical polyp cases. Each case $i$ is represented by a latent scene representation $S_i$ and an associated clinical characterization $y_i \in \{1, 2, \ldots, C\}$, where $C$ is the number of distinct diagnostic categories. Given a new polyp, represented by its latent scene representation $S_{new}$, our objective is to infer its clinical label $y_{new}$. Traditional supervised methods learn a direct mapping $y_i = f_\theta(I_i)$ from images $I_i$ to labels, but these models often lack transparency.

Our EndoFinder framework adopts a transductive approach based on the hypothesis that \textit{polyps with similar latent scene representations likely share common clinical characteristics}. Instead of learning a direct mapping, EndoFinder generates a scene representation $S_{new}$ for a query polyp from its multi-view images. It then retrieves the most similar historical cases from the database $D$. The label for the new polyp is determined by a majority vote among the labels of its $K$ nearest neighbors. Formally, the predicted label $y_{new}$ is given by:

\begin{equation}
y_{new} = \underset{c \in \{1, \dots, C\}}{\mathrm{argmax}} \sum_{k \in \mathcal{N}(S_{new})} \mathbf{1}_{\{y_k = c\}}
\end{equation}
where $\mathcal{N}(S_{new})$ is the set of indices of the $K$ nearest neighbors to $S_{new}$ in the database, and $\mathbf{1}_{\{\cdot\}}$ is the indicator function. This retrieval-based mechanism provides an inherently interpretable diagnosis by referencing similar, previously diagnosed cases.

\subsection{Framework Overview} 
Figure~\ref{fig:workflow} illustrates the EndoFinder architecture. The core objective is to construct a semantically meaningful and retrieval-friendly scene representation space for polyps. The process begins with multi-view endoscopic images $\{I_1, I_2, \dots, I_n\}$ of a target polyp. An image encoder $E_\phi$ extracts feature embeddings from each view. These embeddings are then fused by a scene encoder into a unified latent scene representation $S = E_\theta(E_\phi(I_1), \dots, E_\phi(I_n))$, which encapsulates the polyp's 3D-aware appearance. For rapid retrieval, this continuous representation is converted into a binary hash code. During inference, the query polyp's hash code is compared against the reference database using Hamming distance. A KNN search retrieves similar cases, whose associated diagnoses and images support the clinical decision-making process.

\subsection{Self-supervised Learning for Single-View Representation}
\label{sec:one-view}

To obtain robust single-view polyp embeddings, we pre-train a ViT-based image encoder by combining generative and contrastive self-supervised learning. For the contrastive branch, we follow a SimCLR-based strategy where augmented views of the same image are positive pairs. Inspired by SSCD~\citep{pizzi2022self}, we use its advanced augmentation strategy (including MixUp and CutMix) and a combined loss of InfoNCE and entropy regularization to improve feature discrimination. For the generative branch, we incorporate a masked autoencoder (MAE) objective, where the decoder is trained to reconstruct randomly masked image patches. Crucially, to address the low inter-class variance between polyps and surrounding tissue, we use polyp segmentation masks to guide the masking process, preserving informative polyp regions while masking the background. This forces the model to focus on clinically significant areas.

The training pipeline for single-view representation learning is shown in Figure~\ref{fig:one-view}. For each input image, two augmented views are generated. An optional polyp segmentation mask is applied to retain only the lesion region. A learnable [CLS] token is prepended to the sequence of patch embeddings, which is then fed into the ViT encoder. The output [CLS] embeddings are used for contrastive learning, while the output patch embeddings corresponding to the masked input are passed to a decoder for reconstruction.

The contrastive loss consists of the InfoNCE loss and an entropy loss. For a batch of $N$ images, yielding $2N$ augmented views, the positive pairs are $P = \{(i, i+N), (i+N, i)\}_{i \in \{1, \dots, N\}}$. Let $P_i$ be the matching set for image $i$. The temperature-scaled cosine similarity between representations $z_i$ and $z_j$ is $s_{i, j} = z_i^T z_j / \tau$. The pairwise loss is:
\begin{equation}
\ell_{i,j} = -\log \frac{\exp(s_{i,j})}{\exp(s_{i,j}) + \sum_{k \notin \hat{P_i}} \exp(s_{i,k})}
\end{equation}
where $\hat{P_i} = P_i \cup \{i\}$. The InfoNCE loss is the average over all positive pairs:
\begin{equation}
\mathcal{L}_{InfoNCE} = \frac{1}{2N} \sum_{i=1}^{2N} \frac{1}{|P_i|} \sum_{j \in P_i} \ell_{i,j}
\end{equation}

To further enhance negative sample separation, we add an entropy loss:
\begin{equation}
\label{equ:entropy loss}
\mathcal{L}_{entropy} = -\frac{1}{2N} \sum_{i=1}^{2N} \log \left( \min_{j \notin \hat{P_i}} \| z_i - z_j \|_2 \right)
\end{equation}

The total contrastive objective is a weighted sum:
\begin{equation}
\label{equ:CON loss}
\mathcal{L}_{CON} = \mathcal{L}_{InfoNCE} + \lambda \mathcal{L}_{entropy}
\end{equation}

For the generative branch, the decoder reconstructs only the masked image portions using Mean Squared Error (MSE):
\begin{equation}
\label{equ:MSE Loss}
\mathcal{L}_{MSE} = \frac{1}{2N} \sum_{i=1}^{2N} \frac{1}{|M_i|} \sum_{k \in M_i} (\hat{I}_{i,k} - I_{i,k})^2
\end{equation}
where $M_i$ is the set of masked pixel indices for image $i$, and $\hat{I}_{i,k}$ and $I_{i,k}$ are the reconstructed and original pixel values.

The final training objective combines both losses:
\begin{equation}
\label{equ:Loss}
\mathcal{L} = \mathcal{L}_{CON} + \gamma \mathcal{L}_{MSE}
\end{equation}
where $\lambda$ and $\gamma$ are hyperparameters.

\subsection{Multi-view Latent Scene Representation}

\begin{figure*}[h]
\centering
\includegraphics[scale=.25]{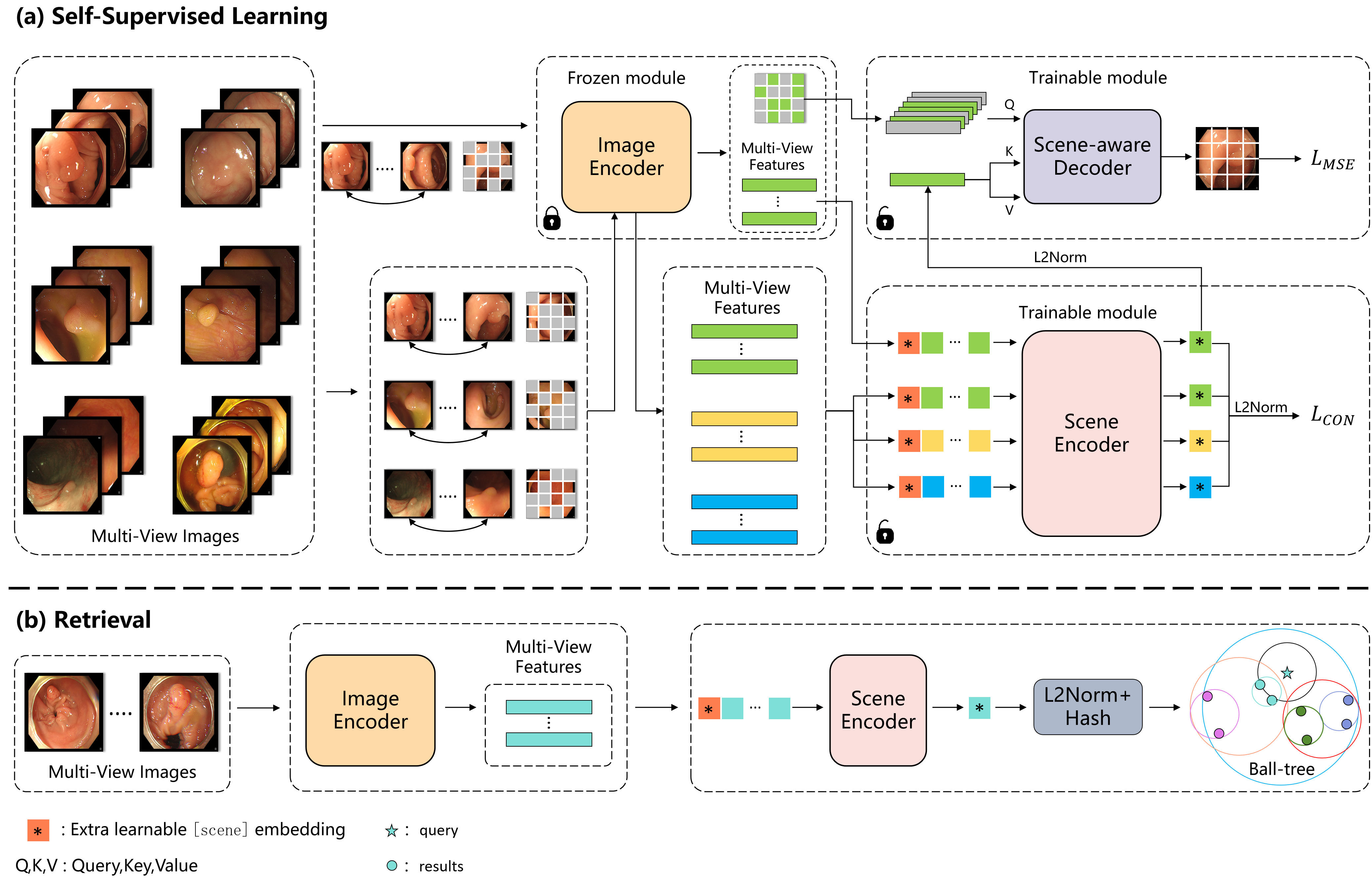}
\caption{Overview of the multi-view scene representation learning framework. 
Polyp images from multiple views are encoded individually via a frozen image encoder $E_\phi$, then fused through a scene encoder $E_\theta$ using self-attention. 
A learnable [Scene] token aggregates scene-level information, which is supervised by contrastive loss across views and a cross-attention-based reconstruction loss.
}
\label{fig:multi-view}
\end{figure*}
To obtain a comprehensive representation of polyps, we introduce a latent scene representation framework that integrates multiple views into a unified scene-level embedding. This approach is more robust to viewpoint variations and occlusions than single-view methods. Given a pre-trained image encoder $E_\phi$ (from Section~\ref{sec:one-view}), we train a scene encoder $E_\theta$ to fuse the embeddings of $n$ views $\{I_1, \dots, I_n\}$ of the same polyp into a latent representation $S = E_\theta(E_\phi(I_1), \dots, E_\phi(I_n))$.

\vspace{0.5em}
\noindent\textbf{Contrastive learning.}
During training, for each scene with $n$ views, we generate $n$ combinations by selecting one view for reconstruction and using the remaining $n-1$ views to compute the scene token. Combinations from the same polyp are treated as mutual positives. Let $P$ be the set of all positive scene-level pairs $(i, j)$ in a batch. Given the temperature-scaled cosine similarity $s_{i,j} = S_i^T S_j / \tau$, the scene-level InfoNCE loss is:

Given the temperature-scaled cosine similarity $s_{i,j} = S_i^T S_j / \tau$, the InfoNCE loss is computed as:

\begin{equation}
\mathcal{L}_{InfoNCE}^{scene} =  \frac{1}{|P|}\sum\limits_{{i,j}\in{P}}-\log\frac{\exp(s_{i,j})}{\sum_{k\neq i}\exp(s_{i,k})}
\end{equation}

We also apply entropy regularization to encourage a uniform distribution of representations:

\begin{equation}
\mathcal{L}_{entropy}^{scene} = -\frac{1}{N}
\sum_{i=1}^{N}
\log{\left(\min_{j \notin \hat{P_i}} \|S_i - S_j\|_2 \right)}
\label{equ:entropy-scene}
\end{equation}
where $N$ is the total number of combinations in the batch and $\hat{P_i}$ is the set of positive partners for scene $i$ including itself.

\vspace{0.5em}
\noindent\textbf{Cross-attention reconstruction.}
To enrich the scene embedding with spatial semantics, we adopt a masked reconstruction objective inspired by RUST~\citep{sajjadi2023rust}. One image per scene is randomly selected for partial reconstruction, while the remaining views generate the latent scene representation.

Let $z_t \in \mathbb{R}^{L \times D}$ be the visible token features from the masked image, and $S \in \mathbb{R}^{1 \times D}$ be the scene representation. Reconstruction is performed using cross-attention between the scene representation (as key) and the visible tokens (as query). The decoder then reconstructs the masked patches, and we apply the reconstruction loss:

\begin{equation}
\label{equ:MSE Loss scene}
\mathcal{L}_{MSE}^{scene} = \frac{1}{N}\sum_{i=1}^{N}\frac{1}{|M_i|}\sum_{k \in M_i} (\hat{I}_{i,k} - I_{i,k})^2
\end{equation}

\paragraph{Overall objective.}
The total loss for training the scene encoder combines the contrastive and reconstruction losses:

\begin{equation}
\mathcal{L}_{total}^{scene} = \mathcal{L}_{InfoNCE}^{scene} + \lambda_{ent} \mathcal{L}_{entropy}^{scene} + \lambda_{rec} \mathcal{L}_{MSE}^{scene}
\end{equation}
where $\lambda_{ent}$ and $\lambda_{rec}$ are weighting hyperparameters.

\vspace{0.5em}
\noindent\textbf{Training strategy.}
The training procedure is detailed in Algorithm~\ref{alg:training_procedure}. Given a set of $n$ views $\{v_1, \dots, v_n\}$ from a polyp scene, we use $n-1$ views to construct the latent scene representation. The final view $v_n$ is reserved for masked reconstruction. We jointly optimize a contrastive loss for scene-level consistency, an entropy regularization term, and a reconstruction loss on the masked tokens from $v_n$.

\begin{algorithm}
\caption{Training Procedure of Our Method}
\label{alg:training_procedure}
\textbf{Input:} Image Encoder $E_\phi$, PolypScene-2k Data $\mathcal{D}$ \\
\textbf{Output:} Scene Encoder $E_\theta$, Scene-aware Decoder $D_\theta$
\begin{algorithmic}[1]
\For{$e \in \{1, \dots, E\}$}
    \For{Mini-batch $B$ in $\mathcal{D}$}
        \State $z$, $tokens$ = $E_\phi(B)$ \Comment{Extract single-view features}
        \State $S$ = $E_\theta(z_{v_1}, \dots, z_{v_{n-1}})$ \Comment{Fuse multi-view features}
        \State Compute contrastive loss $\mathcal{L}_{InfoNCE}^{scene} + \lambda_{ent} \cdot \mathcal{L}_{entropy}^{scene}$ 
        \State $\hat{tokens}_{v_n}$ = \texttt{mask}($tokens_{v_n}$)
        \State $\hat{I}$ = $D_\theta(S, \hat{tokens}_{v_n})$ \Comment{Reconstruct masked view}
        \State Compute reconstruction loss $\mathcal{L}_{MSE}^{scene}$
        \State Update $E_\theta$, $D_\theta$ by minimizing total loss $\mathcal{L}_{total}^{scene}$
    \EndFor
\EndFor
\end{algorithmic}
\end{algorithm}

\subsection{Hash-based Retrieval}
To enable efficient similarity search, we transform each latent scene representation $S$ into a compact binary hash code using a sign-based quantization function:

\begin{equation}
\bar{S}_{s,k} = 
\begin{cases} 
1 & \text{if } S_{s,k} \geq 0, \\
-1 & \text{if } S_{s,k} < 0.
\end{cases}
\end{equation}
Here, $S_{s,k}$ is the $k$-th dimension of the latent scene embedding for scene $s$, and $\bar{S}_{s,k}$ is its binarized value. The final hash code lies in $\{-1, 1\}^K$, where $K$ is the hash length.

At inference time, we compute the binary hash code for a query polyp and retrieve its $k$-nearest neighbors from a pre-compiled database using Hamming distance. To accelerate this process, we build a ball tree index over the binary codes. A majority vote among the retrieved neighbors' labels determines the final diagnosis for the query case. This retrieval-based prediction is not only interpretable but also robust, as it leverages collective evidence from multiple similar examples.

\subsection{Implementation Details}
All models were trained on an NVIDIA GeForce RTX 3090 GPU. We used ViT-L/16 as our Image Encoder and a standard MAE configuration for its decoder. The Scene Encoder was a one-layer Transformer, and the Scene-aware Decoder was a two-layer Transformer with cross-attention. First, the Image Encoder was trained on the Polyp-18k dataset for 100 epochs. We used SSCD's advanced+mixup augmentation, a 50\% masking ratio, and polyp masks to preserve lesion regions. We used the AdamW optimizer with a learning rate of 1.25e-4, a weight decay of 0.05, and a cosine annealing scheduler. The contrastive learning temperature $\tau$ was set to 0.05. The effective batch size was 64. Next, with the Image Encoder frozen, the Scene Encoder was trained on the PolypScene-2k dataset for 200 epochs. The input to the Scene Encoder was the global [CLS] token representation from the Image Encoder for each view. We used SSCD's advanced augmentation, a batch size of 512, and an AdamW optimizer with a learning rate of 1e-4. This strategy allowed us to learn an intelligent feature fusion method while retaining the powerful single-view representations.

\section{Datasets and Experimental Design}
Our experimental evaluation is structured across several progressive tasks, utilizing a suite of in-house datasets (summarized in Table~\ref{tab:dataset_summary}).

\begin{table*}[ht]
\centering
\caption{Summary of Datasets Used in This Study.}
\label{tab:dataset_summary}
\begin{tabular}{|l|c|c|l|}
\hline
\textbf{Dataset Name} & \textbf{\#Polyps} & \textbf{Views per Polyp} & \textbf{Purpose} \\ \hline
Polyp-18k & 17,969 & 1 & Image encoder pre-training \\ \hline
Polyp-Twin & 200 & 2 & Single-view re-identification \\ \hline
PolypScene-2k & 2,219 & 4 & Scene representation learning \\ \hline
PolypScene-250 & 250 & 4 & Scene-based re-identification \\ \hline
PolypScene-80 & 80 & 4 & Pathological classification \\ \hline
\end{tabular}
\end{table*}

First, to establish a robust feature extractor, we pre-trained our Polyp-aware Image Encoder on \textbf{Polyp-18k}, a large-scale collection of 17,969 polyp images with corresponding segmentation masks. The encoder's ability to generate consistent representations from single viewpoints was then benchmarked on a single-view re-identification task using the \textbf{Polyp-Twin} dataset, which contains 200 polyps, each captured in two distinct video frames. Having established a strong image encoder, we proceeded to learn multi-view scene representations. For this, we trained the Scene Representation Transformer on \textbf{PolypScene-2k}, a large multi-view dataset containing 2,219 polyps, each imaged from four different angles. The discriminative power of the resulting scene embeddings was then evaluated on a scene-based re-identification task using the \textbf{PolypScene-250} benchmark, which features 250 polyps with four views each. To assess the clinical utility of our framework, we performed a pathological classification task on \textbf{PolypScene-80}, a subset of PolypScene-250 containing 80 polyps with pathologically confirmed binary labels (34 vs. 46). Finally, to confirm the feasibility of real-time application, we evaluated the performance of hash-based retrieval on the PolypScene-250 dataset, analyzing the trade-off between retrieval speed and accuracy.

    
    
    
    

\section{Results}

\subsection{Performance on Polyp Re-Identification}

\begin{figure*}[!h]
\centering
\includegraphics[scale=.4]{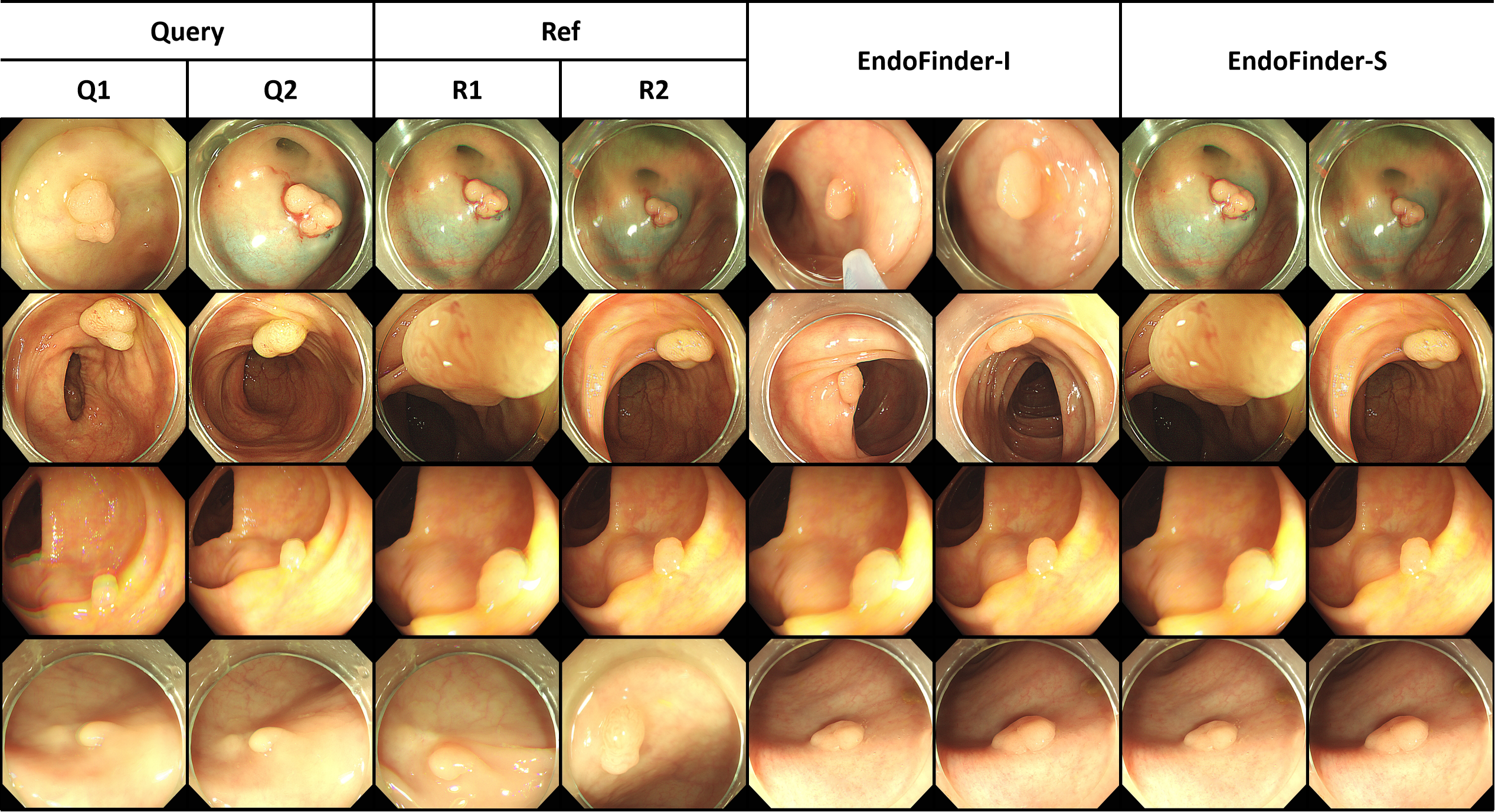}
\caption{Qualitative comparison of re-identification performance. The Query set contains two views (Q1, Q2) of a polyp. The Ref set shows the correct match (R1, R2). EndoFinder-I uses averaged features, while EndoFinder-S uses the Scene Transformer. The Scene Transformer correctly identifies the polyp, whereas the simpler averaging method fails.}
\label{fig:E2}
\end{figure*}

\subsubsection{Single-view re-identification}
We first evaluated single-view polyp re-identification on the Polyp-twin dataset. We used Mean Average Precision ($\mu$AP), Top-1 Accuracy (Acc@1), and Recall at 90\% Precision (Recall@P90) as metrics. As shown in Table \ref{tab:one_to_one}, our Image Encoder (EndoFinder-I) outperformed other leading self-supervised methods. However, the overall scores highlight the inherent difficulty of re-identifying a polyp from a single, potentially ambiguous viewpoint, motivating our multi-view approach.

\begin{table}[h]
\caption{Performance of different methods in single view re-identification tasks}
\label{tab:one_to_one}
\begin{tabular}{llccc}
\hline
\multicolumn{1}{l}{Method} & Backbone & $\mu AP$ & $Acc@1$ & $Recall@P90$ \\ \hline
Dino                       & ViT-L    & 0.52   & 0.59  & 0.30       \\
SSCD                       & Resnet50 & 0.64   & 0.69  & 0.53       \\
MAE                        & ViT-L    & 0.45   & 0.52  & 0.29       \\ 
EndoFinder-I               & ViT-L    & \textbf{0.67}   & \textbf{0.70}  & \textbf{0.56}       \\ \hline
\end{tabular}
\end{table}

\subsubsection{Multi-view re-identification}
To validate our multi-view approach, we conducted experiments on the PolypScene-250 dataset. First, using only the pre-trained Image Encoder (EndoFinder-I), we averaged the features of multiple views to form a single representation. 

We gradually increased the number of views used for queries and references, starting from single-view retrieval and extending to cases where two views were used for both queries and references. This allowed us to comprehensively evaluate recognition performance under different retrieval settings. Each polyp in the PolypScence dataset has four views, which are randomly labeled as Q1, Q2, R1, and R2. Views Q1 and Q2 are assigned to the query group, while views R1 and R2 are assigned to the reference (ref) group.
\begin{itemize}
    \item If both Q2 and R1 have a $\surd$ mark, it represents a single-view Re-ID involving two views.
    \item If Q2, R1, and R2 have $\surd$ marks, it indicates a single-view-to-multi-view Re-ID.
    \item If both Q1 and Q2 are marked, while only R1 is marked in the ref group, it signifies a multi-view-to-single-view Re-ID.
    \item If all four views have $\surd$ marks, it represents a multi-view-to-multi-view Re-ID scenario.
\end{itemize}

As shown in Table \ref{tab:impact_add_views}, increasing the number of views in either the query or the reference set consistently improved retrieval performance. The best results with this averaging strategy were achieved in the two-views-vs-two-views setting, confirming that aggregating information from multiple perspectives enhances robustness.

\begin{table}[h]
\centering
\caption{Impact of increasing views on re-identification performance.}
\label{tab:impact_add_views}
\begin{tabular}{cc cc ccc}
\hline
\multicolumn{2}{c}{Query Views} & \multicolumn{2}{c}{Ref Views} & \multicolumn{3}{c}{Performance} \\
\cline{1-2} \cline{3-4} \cline{5-7}
Q1 & Q2 & R1 & R2 & $\mu AP$ & $Acc@1$ & $Recall@P90$ \\
\hline
$\surd$ &       & $\surd$ &       & 0.27 & 0.35 & 0.13 \\
$\surd$ &       &         & $\surd$ & 0.49 & 0.53 & 0.41 \\
       & $\surd$ & $\surd$ &       & 0.34 & 0.41 & 0.20 \\
       & $\surd$ &         & $\surd$ & 0.60 & 0.64 & 0.52 \\
$\surd$ &       & $\surd$ & $\surd$ & 0.50 & 0.55 & 0.39 \\
       & $\surd$ & $\surd$ & $\surd$ & 0.60 & 0.64 & 0.48 \\
$\surd$ & $\surd$ & $\surd$ &       & 0.38 & 0.45 & 0.26 \\
$\surd$ & $\surd$ &         & $\surd$ & 0.64 & 0.66 & \textbf{0.59} \\
$\surd$ & $\surd$ & $\surd$ & $\surd$ & \textbf{0.68} & \textbf{0.70} & 0.58 \\
\hline
\end{tabular}
\end{table}

\subsubsection{Scene-based re-identification}
Next, we evaluated the performance of our trained Scene Encoder (EndoFinder-S), which provides a more sophisticated fusion of multi-view features. We compared it against the simple averaging baseline (EndoFinder-I) and other methods where a scene encoder was trained on top of their respective pre-trained image encoders. The results, shown in Table \ref{tab:ReID}, demonstrate that EndoFinder-S achieves the best performance, surpassing all other methods, including its own averaging-based counterpart. This confirms that our learned scene representation effectively integrates multi-view information. Notably, methods based on the Transformer architecture (DINO, MAE) also benefited from the scene encoder, indicating the generalizability of our approach. However, the CNN-based SSCD saw a performance decrease, suggesting that our scene encoder is better optimized for ViT-based features. Figure \ref{fig:E2} provides a qualitative example where the scene encoder succeeds in a difficult retrieval case where the simpler averaging method fails.

\begin{table}[h]
\caption{Multi-view re-identification performance on PolypScene-250. The top section shows results for average feature fusion, while the bottom section shows results for Scene Transformer fusion.}
\label{tab:ReID}
\begin{tabular}{clccc}
\hline
\multicolumn{1}{l}{}                                                          & \multicolumn{1}{l}{Method} & $\mu AP$      & $Acc@1$       & $Recall@P90$  \\ \hline
\multirow{4}{*}{\begin{tabular}[c]{@{}c@{}}Average\\ fusion\end{tabular}}     & Dino                       & 0.55          & 0.63          & 0.33          \\
                                                                              & SSCD                       & 0.63          & 0.66          & 0.58          \\
                                                                              & MAE                        & 0.58          & 0.64          & 0.43          \\
                                                                              & EndoFinder-I                 & 0.68          & 0.70          & 0.58          \\ \hline
\multirow{4}{*}{\begin{tabular}[c]{@{}c@{}}Transformer\\ fusion\end{tabular}} & Dino-S                       & 0.65          & 0.70          & 0.54          \\
                                                                              & SSCD-S                       & 0.60          & 0.64          & 0.48          \\
                                                                              & MAE-S                        & 0.58          & 0.64          & 0.42          \\
                                                                              & EndoFinder-S                 & \textbf{0.71} & \textbf{0.74} & \textbf{0.59} \\ \hline
\end{tabular}
\end{table}

\subsection{Application on Optical Polyp Diagnosis}
We evaluated the clinical relevance of our learned scene representations by performing pathology classification on the PolypScene-80 dataset. Using five-fold cross-validation, we employed a k-NN classifier on the scene embeddings, where the majority vote of the retrieved neighbors determines the diagnosis. We compared our full method (EndoFinder-S) against the averaging baseline (EndoFinder-I) and standard supervised classifiers fine-tuned on the same data.

\begin{figure*}[h]
\centering
\includegraphics[scale=.40]{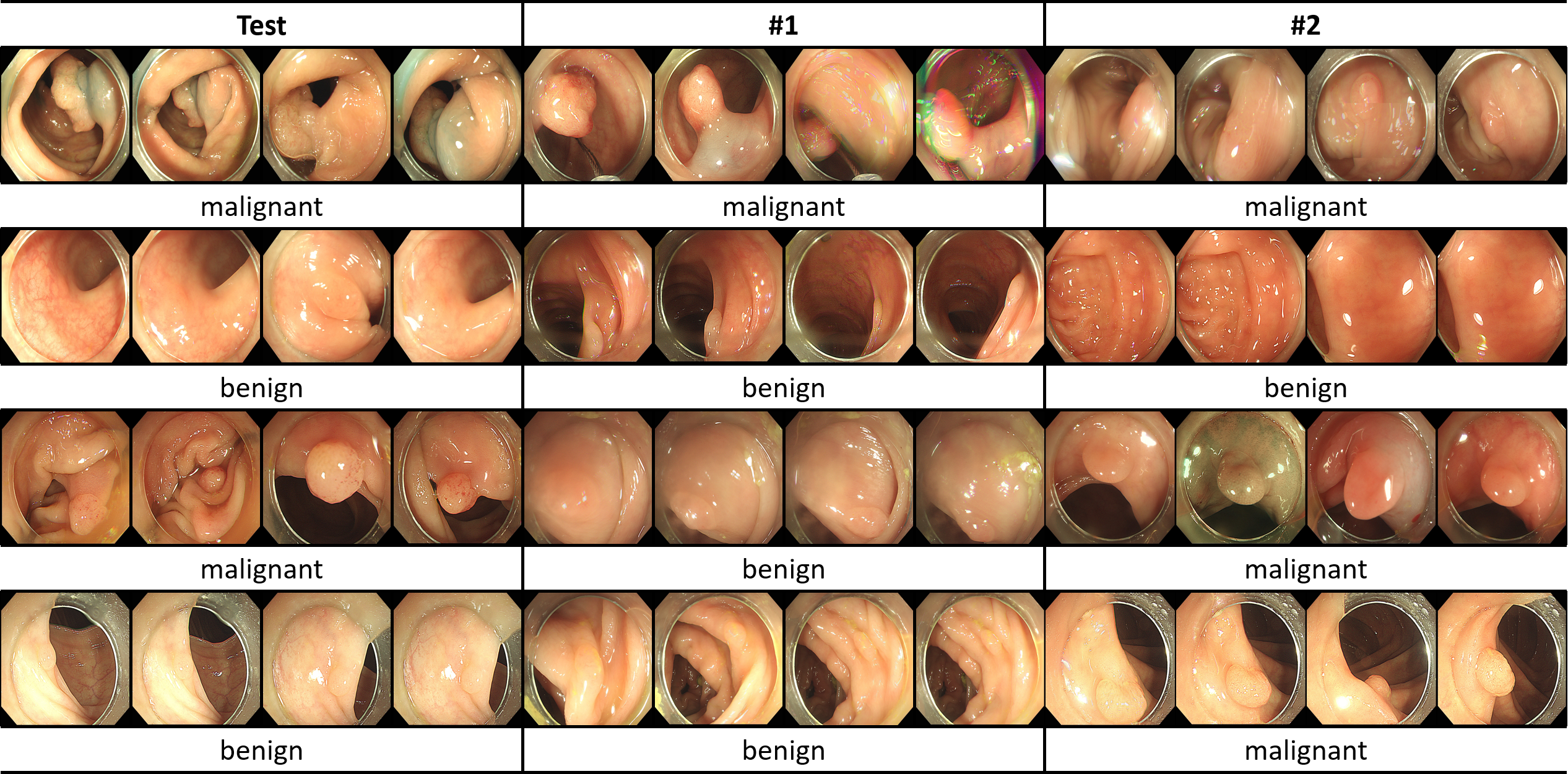}
\caption{Example of nearest neighbor classification using EndoFinder-S. The table is divided into three column groups: Query, \#1, and \#2. Each group contains four images. The Query column represents the data awaiting classification, with four different views for each case. The \#1 column shows the closest retrieval result, while the \#2 column presents the second most similar retrieval result. A total of four polyps are awaiting classification, and their deterioration levels are annotated below each of the four views.}
\label{fig:E3}
\end{figure*}

As shown in Table \ref{tab:cls}, EndoFinder-S with k-NN (k=6) achieved the best performance, outperforming all other methods across AUC, Accuracy, and F1-score. This result is significant because our retrieval-based method not only provides superior accuracy but also offers inherent interpretability by presenting clinicians with visually similar, pathologically-confirmed reference cases. Figure \ref{fig:E3} shows qualitative examples of this nearest-neighbor classification, demonstrating how the system retrieves relevant cases to support its diagnosis.

\begin{table}[h]
\caption{Pathology classification performance on the PolypScene-80 dataset using five-fold cross-validation. Our retrieval-based method (EndoFinder-S) outperforms standard supervised models.}

\label{tab:cls}
\begin{tabular}{clccc}
\hline
                                                                          & Method      & \multicolumn{1}{c}{AUC} & \multicolumn{1}{c}{ACC} & \multicolumn{1}{c}{F1} \\ \hline
\multirow{3}{*}{\begin{tabular}[c]{@{}c@{}}Fine\\ tuning\end{tabular}}    & Resnet50    & 77.33                   & 73.75                   & 77.97                  \\
                                                                          & Densenet121 & 76.66                   & 70.00                   & 72.68                  \\
                                                                          & ViT-L       & 81.74                   & \textbf{78.75}          & 80.08                  \\ \hline
\multirow{4}{*}{\begin{tabular}[c]{@{}c@{}}Average\\ fusion\end{tabular}} & MAE         & 61.48                   & 62.50                    & 70.27                  \\
                                                                          & Dino        & 74.41                   & 75.00                    & 80.13                  \\
                                                                          & SSCD        & 81.69                   & 76.25                   & 77.63                  \\
                                                                          & EndoFinder-I  & 81.75                   & 67.50                    & 65.48                  \\ \hline
\begin{tabular}[c]{@{}c@{}}Transformer\\ fusion\end{tabular}              & EndoFinder-S  & \textbf{85.59}          & \textbf{78.75}          & \textbf{81.00}         \\ \hline
\end{tabular}
\end{table}

\subsection{Hash Encoding for Efficient Retrieval}

\begin{figure}[h]
\centering
\includegraphics[scale=.38]{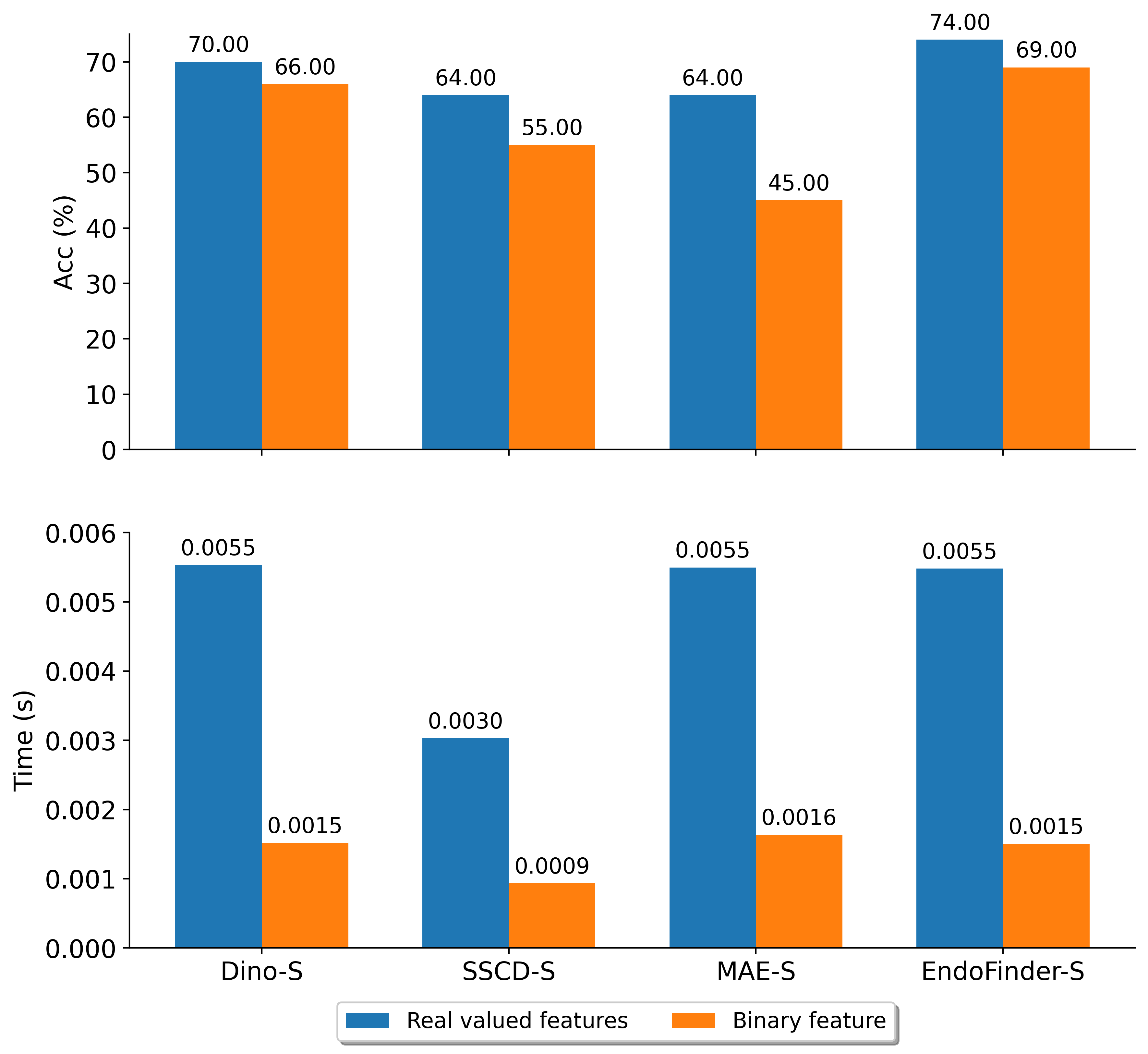}
\caption{The speed improvement and accuracy change brought by using Hamming distance to compare cosine similarity when using ball tree retrieval. Compare the performance of four different scene Transformer-based feature fusion methods in the Re-ID task after applying hash techniques.}
\label{fig:hash}
\end{figure}

To ensure real-time performance, we converted the scene representations into binary hash codes and used a ball tree with Hamming distance for retrieval. On the PolypScene-250 dataset, this approach yielded a significant speedup compared to using cosine similarity on the original float-valued features. As shown in Figure \ref{fig:hash}, our method (EndoFinder-S) achieved a 4x speed improvement with minor drop in $\mu$AP. This trade-off is highly favorable for clinical application, where speed is critical. The accuracy, even after hashing, remains superior to the simpler average feature fusion baseline, demonstrating the robustness of the learned representations.

\section{Discussion} 
In this work, we introduced EndoFinder, a retrieval-based framework for explainable polyp diagnosis that leverages multi-view scene representations. Our approach addresses two critical limitations of current AI systems in medicine: the lack of interpretability and the heavy reliance on large, labeled datasets.

\noindent\textbf{Overall performance.}
The results from the Polyp-Twin dataset demonstrate that our EndoFinder framework exhibits superior image re-identification capability, validating that our method effectively learns more informative polyp representations, as shown in Table \ref{tab:one_to_one}. Expanding to the PolypScene-250 dataset, we first extract multi-view features of polyps using our image encoder and then apply an average fusion strategy. To explore the advantages of multi-view information, we compare performance across different numbers of query and reference images. As shown in Table \ref{tab:impact_add_views}, the best re-identification performance is achieved when both the query and reference sets contain two images from different viewpoints. Furthermore, adding additional images to either the query or reference set consistently improves re-identification accuracy.

As shown in Table \ref{tab:ReID}, when we employ our scene encoder to integrate the features extracted by the image encoder, our method and DINO outperform the average feature fusion approach. However, MAE exhibits only a marginal difference, likely because MAE learns general image features but lacks explicit re-identification guidance, relying instead on holistic image representations for retrieval. Interestingly, the SSCD method, which employs a CNN-based image encoder, experiences a decline in performance after incorporating the scene encoder. This suggests that our scene encoder requires further adaptation to effectively process CNN-derived features. When incorporating hash-based techniques, our method and DINO achieve a 4× speedup at the cost of 6.7\% and 5.7\% accuracy loss, respectively, as illustrated in Figure \ref{fig:hash}. Notably, for methods with weaker baseline performance, hashing results in even greater accuracy degradation, as evidenced by the other two methods in the figure.

For polyp diagnosis, evaluated on the PolypScene-80 dataset, our approach outperforms both the kNN-based average feature fusion method and conventional supervised classifiers, as shown in Table \ref{tab:cls}. It is worth noting that validation was conducted using five-fold cross-validation. Additionally, in the kNN classification setup, four-fifths of the data were treated as a reference database without retraining, whereas the supervised classifier was fine-tuned using pre-trained weights. In summary, our method achieves state-of-the-art performance in both re-identification and diagnostic tasks.

\noindent\textbf{Image retrieval and Embeddings.}
As artificial intelligence and digital medicine become increasingly integrated into healthcare, a robust governance framework is essential to ensure ethical, safe, and effective implementation. Medical image retrieval has emerged as a key component of clinical data management, playing a crucial role in decision-making and patient information protection~\citep{nan2025revisiting}. Traditional supervised classification methods encode knowledge into model weights, which contributes to their “black-box” nature and poses significant challenges to interpretability. In contrast, image retrieval relies on feature representations to retrieve similar samples, where knowledge is typically stored in the form of image embeddings and associated labels. This approach offers several advantages in terms of data protection, transparency, interpretability, avoiding retraining, simplifying validation and testing, and enabling human oversight.

Combining a kNN classifier with embeddings for polyp diagnosis enhances these characteristics~\citep{doerrich2024integrating}. By retrieving and ranking the most similar samples, the final diagnosis is determined through a voting mechanism based on the labels of the retrieved cases. Since these reference samples include optical images, the decision process gains inherent visual interpretability. Additionally, storing embeddings alongside labels in a database not only serves as an effective data management strategy but also simplifies testing and validation. As the dataset scales, the model does not require retraining and can immediately leverage newly added data. Conversely, if certain data points need to be removed due to privacy concerns, they can simply be deleted from the database, ensuring that the corresponding knowledge is also eliminated without necessitating model retraining. This flexible and dynamic knowledge management system makes image retrieval a compelling alternative to traditional supervised classification in medical diagnostics.

\noindent\textbf{Scene Representation Assessment.}
During colonoscopy, physicians naturally inspect polyps from multiple angles. This variation in viewpoint, while challenging for single-view models, provides a wealth of 3D information~\citep{cai2024know}. Our work embraces this clinical reality by treating each polyp as a "scene." By explicitly modeling the relationship between different views, our Scene Representation Transformer learns a more holistic and robust embedding that is less susceptible to the appearance changes caused by a single viewpoint. The superior performance of this approach over both single-view and simple multi-view averaging methods validates this core hypothesis.

\noindent\textbf{Limitations and Future Work.}
A limitation of our current framework is its use of polyp segmentation masks to guide the image encoder, as segmentation errors could propagate. Future work could explore weakly-supervised or unsupervised methods to relax this requirement. Another challenge is the need for multiple views, which may not always be captured in routine practice. Developing methods to work with a variable number of views or even synthesize novel viewpoints could broaden the applicability of EndoFinder. Future research could also extend this scene-based retrieval paradigm to other types of lesions and incorporate additional data modalities, such as patient history or electronic health records, to further refine diagnostic predictions.

\section{Conclusion}
This paper introduced EndoFinder, a retrieval-based diagnostic framework that conceptualizes polyps as 3D scenes to learn robust, multi-view representations. By fusing information from different endoscopic viewpoints, our method significantly improves performance in both polyp re-identification and pathological classification tasks. The retrieval-based paradigm offers a crucial layer of interpretability, allowing clinicians to validate an AI-generated diagnosis by examining similar historical cases with known outcomes. This approach fosters trust and aligns more closely with clinical reasoning. Our results demonstrate that moving from single-image analysis to scene-based retrieval is a promising direction for developing more accurate, transparent, and trustworthy AI systems for endoscopic diagnosis.

\section*{Acknowledgments}
This work was supported in part by the National Key Research and Development Program of China (2024YFF1207500), the National Natural Science Foundation of China (82203193), the Shanghai Municipal Education Commission Project for Promoting Research Paradigm Reform and Empowering Disciplinary Advancement through Artificial Intelligence (SOF101020), and the International Science and Technology Cooperation Program under the 2023 Shanghai Action Plan for Science (23410710400). The computations were performed using the CFFF platform of Fudan University.

\section*{Author contributions}
R-J. Yang developed the methodology, conducted experiments, prepared tables and figures, and drafted the manuscript. Y. Zhu designed the overall study, contributed to data analysis, and assisted in data collection and preparation. P-Y. Fu and P-H. Zhou contributed to data collection and preparation. Q-L. Li contributed to data annotation. Y-Z. Zhang provided technical support during the experiments. Z-H. Wang provided methodological advice and assisted in interpreting results. X. Yang and S. Wang provided guidance on experimental design and contributed to manuscript drafting and revision. All authors read and approved the final manuscript.

\bibliographystyle{model2-names.bst}\biboptions{authoryear}
\bibliography{refs}



\end{document}